\documentclass[aps,onecolumn]{revtex4}
\usepackage{graphicx}
\usepackage{epsfig}
\usepackage{epstopdf}
\usepackage{amsfonts}
\usepackage{amssymb}
\usepackage{amsbsy}
\usepackage{amsmath}
\usepackage{mathrsfs}
\usepackage{latexsym}
\usepackage{natbib}
\usepackage{bm}
\usepackage{color}
\usepackage[a4paper, total={6.4in, 9.8in}]{geometry}

\newcommand{\notimplies}{%
  \mathrel{{\ooalign{\hidewidth$\not\phantom{=}$\hidewidth\cr$\implies$}}}}

\newtheorem{theorem}{Theorem}
\usepackage{braket}
\usepackage{slashed}
\usepackage{pgfplots}
\numberwithin{equation}{section}

\usepackage{tikz}
\usetikzlibrary{shapes,arrows,shadows}

\begin{document}

\title{Classification of quantum states based on the null energy condition}

\author{Susobhan Mandal}
\email{sm17rs045@iiserkol.ac.in}

\affiliation{ Department of Physical Sciences,\\ 
Indian Institute of Science Education and Research Kolkata,\\
Mohanpur - 741 246, WB, India }


\begin{abstract}
Violation of the null energy condition plays an important role both in the general theory of relativity and quantum field theory in curved spacetimes. Over the years, it has been shown that the violation of the null energy condition leads to instability and violation of causality. In quantum field theory, violation of the energy condition also depends on the quantum states apart from the geometry of curved spacetime. Hence, the quantum effects play an important role in the violation of the null energy condition. We show that the set of all the coherent states does not violate the null energy condition. Further, we also show that under certain conditions, the null energy condition is violated either by a pure state or a mixed state. Furthermore, the dynamical violation of the null energy condition by the quantum states is also discussed here.
\end{abstract}

\maketitle

\section{Introduction}
The singularity theorems \cite{PhysRevLett.14.57, senovilla1998singularity, ford2003classical} are some of the important aspects of general relativity. These have profound consequences in black hole mechanics, also known as black hole thermodynamics (BH). The singularity theorems are proved using the earlier work of Raychaudhuri \cite{PhysRev.98.1123} in the causal structure of geodesic congruences, generalized in  \cite{abreu2011some}. However, proofs of singularity theorems in BH strongly depend on the null energy condition (see \cite{mandal2018revisiting, chrusciel2001regularity, kar2007raychaudhuri}). The null energy condition essentially states that the energy-momentum tensor $T_{\mu\nu}$ of the matter satisfies the following condition
\begin{equation}\label{NEC}
T_{\mu\nu}(x)n^{\mu}(x)n^{\nu}(x)\geq0,
\end{equation}
for any null vector $n^{\mu}$, satisfying $g_{\mu\nu}(x)n^{\mu}(x)n^{\nu}(x)=0$ where $g_{\mu\nu}$ is the metric tensor of the spacetime manifold. However, over the years, violations of the null energy condition (NEC) \cite{rubakov2014null, bouhmadi2014wormholes, toshmatov2017energy, baccetti2012null, mandal2018revisiting} have been found in several physical situations. Later, it has been proposed that average null energy condition (ANEC) \cite{klinkhammer1991averaged, yurtsever1995averaged, yurtsever1995remarks, PhysRevD.44.403, fewster2007averaged, fewster2003null, kontou2013averaged} is the minimum requirement for the study of BH and it holds for any spacetime, based on the work \cite{jacobson1995thermodynamics}. The average null energy condition states that the integral of projection of the matter stress-energy onto the tangent vector of a null geodesic cannot be negative
\begin{equation}\label{ANEC}
\int_{\gamma}T_{\mu\nu}l^{\mu}l^{\nu}\equiv\int_{\gamma}T_{\mu\nu}(\lambda)l^{\mu}(\lambda)l^{\nu}(\lambda)d\lambda\geq0,
\end{equation}
where $l^{\mu}$ is the tangent vector to the null geodesic $\gamma$ and $\lambda$ is the affine parameter. Later, it has also been shown that ANEC can be violated \cite{visser1995scale, urban2010averaged, freivogel2018smeared}. If two points are connected by a null geodesic $\gamma$ parametrized by $\lambda$, then (\ref{ANEC}) can be expressed as 
\begin{equation}
\int_{\gamma}T_{\mu\nu}l^{\mu}l^{\nu}=\int_{\gamma}T_{\mu\nu}(x)l^{\nu}(x)dx^{\mu}=\int_{\gamma}\mathcal{J}_{\mu}(x)dx^{\mu},
\end{equation}
where $\mathcal{J}_{\mu}(x)=T_{\mu\nu}(x)l^{\nu}(x)$. If there exist closed null geodesics \cite{klinkhammer1992vacuum, sarma2013vacuum}, then the same two points must be connected by two null geodesics, let's say $\gamma_{1},\gamma_{2}$, hence,
\begin{equation}
\int_{\gamma_{1}}T_{\mu\nu}l^{\mu}l^{\nu}-\int_{\gamma_{2}}T_{\mu\nu}l^{\mu}l^{\nu}=\int_{\mathcal{C}}\mathcal{J}_{\mu}(x)dx^{\mu}=\int_{\mathcal{S}}\nabla_{[\mu}\mathcal{J}_{\nu]}(x)dx^{\mu}\wedge dx^{\nu},
\end{equation}
where the boundary of the surface $\mathcal{S}$ is given by the closed null geodesic $\mathcal{C}$. In general, the flux-integral in \textit{r.h.s} is non-zero. However, if it is positive definite or negative-definite, then there may exist null geodesics for which (\ref{ANEC}) is violated which follows from the above equation. 

In quantum field theory in curved spacetime (QFTCS), the stress-energy tensor of a matter is given by a hermitian operator, constructed out of field operators. Hence, both the conditions (\ref{NEC}) and (\ref{ANEC}) depend strongly on the quantum states of matter in the curved spacetime. Unfortunately, under what conditions, the quantum states of matter satisfy either of (\ref{NEC}) or (\ref{ANEC}) are ambiguous yet. Hence, it is important to classify quantum states of matter which hold NEC (\ref{NEC}) since it is stronger than ANEC (\ref{ANEC}). However, it is quite clear that in general the collection of all the quantum states of matter, satisfying NEC do not form a vector space since given a null vector $n^{\mu}$,
\begin{equation}\label{0.1}
\bra{\psi_{1}}\hat{T}_{\mu\nu}\ket{\psi_{1}}n^{\mu}n^{\nu}\geq0, \ \bra{\psi_{2}}\hat{T}_{\mu\nu}\ket{\psi_{2}}n^{\mu}n^{\nu}\geq0\notimplies(\bra{\psi_{1}}a^{*}+\bra{\psi_{2}}b^{*})\hat{T}_{\mu\nu}(a\ket{\psi_{1}}+b\ket{\psi_{2}})n^{\mu}n^{\nu}\geq0,
\end{equation}  
for arbitrary complex numbers $a,b$ where $\ket{\psi_{1}}$ and $\ket{\psi_{2}}$ are two quantum states satisfying NEC.

The aim of the present article is to classify a collection of quantum states of the matter in a generic curved spacetime, satisfying NEC. For the sake of mathematical simplicity, we use a non-interacting real scalar field theory. In order to classify these quantum states, we use coherent state descriptions since we follow the canonical approach to QFTCS \cite{ford2002d3, dewitt1975quantum, fulling1989aspects}. We also briefly discuss the classification of quantum states based on ANEC.

\section{Introduction to the coherent states}
In this section, we briefly review the properties of coherent states as the preliminary material for our later studies. In quantum mechanics, the phase space observables in classical mechanics are mapped to linear hermitian operators defined over the Hilbert space through the map $x\mapsto\hat{x}, \ p\mapsto\hat{p}, \ \{x,p\}=1\mapsto[\hat{x},\hat{p}]=i$. Using the operators $\hat{x},\hat{p}$, creation operator $\hat{a}^{\dagger}$ and annihilation operator $\hat{a}$ can be constructed, satisfying the algebra $[\hat{a},\hat{a}^{\dagger}]=1$. A coherent state $\ket{\alpha}$, also known as Glauber state, is an eigenstate of the annihilation operator $\hat{a}$ with eigenvalue $\alpha\in\mathbb{C}$, defined by $\hat{a}\ket{\alpha}=\alpha\ket{\alpha}$. Now, we define the displacement operator
\begin{equation}
\hat{\mathcal{D}}(\alpha)\equiv e^{\alpha\hat{a}^{\dagger}-\alpha^{*}\hat{a}}=e^{-\frac{1}{2}|\alpha|^{2}}e^{\alpha\hat{a}^{\dagger}}e^{-\alpha^{*}\hat{a}},
\end{equation} 
where BCH formula is used in the second equality. The above definition implies $\hat{\mathcal{D}}^{\dagger}(\alpha)=\hat{\mathcal{D}}(-\alpha)=\hat{\mathcal{D}}^{-1}(\alpha)$. Further, it can be shown that
\begin{equation}\label{3}
\hat{\mathcal{D}}^{\dagger}(\alpha)\hat{a}\hat{\mathcal{D}}(\alpha)=\hat{a}+\alpha, \ \hat{\mathcal{D}}^{\dagger}(\alpha)\hat{a}^{\dagger}\hat{\mathcal{D}}(\alpha)=\hat{a}^{\dagger}+\alpha^{*},
\end{equation}
and $\hat{\mathcal{D}}(\alpha+\beta)=\hat{\mathcal{D}}(\alpha)\hat{\mathcal{D}}(\beta)e^{-i\text{Im}(\alpha\beta^{*})}$. The relation (\ref{3}) essentially implies $\ket{\alpha}=\hat{\mathcal{D}}(\alpha)\ket{0}$, where $\ket{0}$ is the state, annihilated by the annihilation operator. Therefore, the coherent states can be obtained through the action of the displacement operator on the state $\ket{0}$ (see \cite{zhang1990coherent} for more details). As a result, the inner product between two coherent states can be obtained
\begin{equation}\label{inner-product}
\braket{\beta|\alpha} = \bra{0}\hat{\mathcal{D}}^{\dagger}(\beta)\hat{\mathcal{D}}(\alpha)\ket{0}=e^{-\frac{|\alpha-\beta|^{2}}{2}+i\text{Im}(\alpha\beta^{*})}.
\end{equation}

\section{Scalar field theory in a generic curved spacetime}
The action for a minimally coupled real massless scalar field theory in a generic curved spacetime is given by
\begin{equation}\label{action}
\mathcal{S}=-\int\sqrt{-g(x)}d^{4}x \ \frac{1}{2}g^{\mu\nu}(x)\partial_{\mu}\phi(x)\partial_{\nu}\phi(x).
\end{equation}
A complete set of mode solutions $\{f_{j},f_{j}^{*}\}$ of the Klein-Gordon equation can be obtained through the extremization of the above action, with $\{j\}$ being a set of discrete or continuous labels distinguishing the independent solutions. These modes are normalized \cite{ford2002d3} \textit{w.r.t} the following inner product
\begin{equation}
\langle f,g\rangle=-i\int\sqrt{-g(x)}d^{3}x[f^{*}(t,\vec{x})\overleftrightarrow{\partial^{0}}g(t,\vec{x})],
\end{equation}
such that
\begin{equation}
\begin{split}
\langle f_{j},f_{j'}\rangle=\delta_{jj'}, & \ \langle f_{j}^{*},f_{j'}^{*}\rangle=-\delta_{jj'}\\
\langle f_{j}^{*},f_{j'}\rangle & =\langle f_{j},f_{j'}^{*}\rangle=0.
\end{split}
\end{equation}
The above inner product is well-defined since it is time-independent which can be easily checked. The corresponding completeness relation can be written as follows
\begin{equation}
\sum_{j}[f_{j}(t,\vec{x})\partial^{0}f_{j}^{*}(t,\vec{x}')-f_{j}^{*}(t,\vec{x} )\partial^{0}f_{j}(t,\vec{x}')]=-\frac{i}{\sqrt{-g(x)}}\delta^{(3)}(\vec{x}-\vec{x}').
\end{equation}
The field operator $\hat{\phi}(x)$ can be written in terms of the above basis solutions as follows
\begin{equation}\label{3.5}
\hat{\phi}(x)=\sum_{j}[\hat{a}_{j}f_{j}(x)+\hat{a}_{j}^{\dagger}f_{j}^{*}(x)].
\end{equation}
with the following commutation relations
\begin{equation}\label{3.6}
[\hat{a}_{j},\hat{a}_{j'}]=0=[\hat{a}_{j}^{\dagger},\hat{a}_{j'}^{\dagger}], \ 
[\hat{a}_{j},\hat{a}_{j'}^{\dagger}]=\delta_{jj'}.
\end{equation}
The vacuum state $\ket{0}$ is defined by
\begin{equation}
\hat{a}_{j}\ket{0}=0, \ \forall j \ .
\end{equation}
The multi-particle states can be obtained by applying the products of creation operators $\{\hat{a}_{j}^{\dagger}\}$ on the vacuum state with a suitable normalization constant
\begin{equation}
\ket{j_{1}j_{2}\ldots j_{n}}=\mathcal{N}\sum_{\sigma\in S_{n}}\prod_{i=1}^{n}(\hat{a}_{j_{\sigma i}}^{\dagger})^{n_{i}}\ket{0},
\end{equation}
where $\mathcal{N}$ is a normalization factor. In terms of the creation and annihilation operators, the stress-energy tensor operator can be expressed as follows
\begin{equation}
\begin{split}
\hat{T}_{\mu\nu} & =\sum_{i,j}[\mathcal{T}_{\mu\nu}(f_{i},f_{j})\hat{a}_{i}\hat{a}_{j}+\mathcal{T}_{\mu\nu}(f_{i},f_{j}^{*})\hat{a}_{i}\hat{a}_{j}^{\dagger}+\mathcal{T}_{\mu\nu}(f_{i}^{*},f_{j})\hat{a}_{i}^{\dagger}\hat{a}_{j}+\mathcal{
T}_{\mu\nu}(f_{i}^{*},f_{j}^{*})\hat{a}_{i}^{\dagger}\hat{a}_{j}^{\dagger}]\\
\implies :\hat{T}_{\mu\nu}: & =\sum_{i,j}[\mathcal{T}_{\mu\nu}(f_{i},f_{j})\hat{a}_{i}\hat{a}_{j}+\mathcal{T}_{\mu\nu}(f_{i},f_{j}^{*})\hat{a}_{j}^{\dagger}\hat{a}_{i}+\mathcal{T}_{\mu\nu}(f_{i}^{*},f_{j})\hat{a}_{i}^{\dagger}\hat{a}_{j}+\mathcal{
T}_{\mu\nu}(f_{i}^{*},f_{j}^{*})\hat{a}_{i}^{\dagger}\hat{a}_{j}^{\dagger}],
\end{split}
\end{equation}
where $: \ :$ is the normal-ordering operation and $\mathcal{T}_{\mu\nu}(f_{i},f_{j})=\partial_{\mu}f_{i}\partial_{\nu}f_{j}$. Normal-ordering is used to make $\bra{0}:\hat{T}_{\mu\nu}:\ket{0}=0$. Further, we obtain the following relation
\begin{equation}\label{operator}
\begin{split}
:\hat{T}_{\mu\nu}:n^{\mu}n^{\nu} & =\sum_{i,j}\Big[\mathcal{N}(f_{i},f_{j})\hat{a}_{i}\hat{a}_{j}+\mathcal{N}(f_{i},f_{j}^{*})\hat{a}_{j}^{\dagger}\hat{a}_{i}+\mathcal{N}(f_{i}^{*},f_{j})\hat{a}_{i}^{\dagger}\hat{a}_{j}+\mathcal{N}(f_{i}^{*},f_{j}^{*})\hat{a}_{i}^{\dagger}\hat{a}_{j}^{\dagger}\Big],
\end{split}
\end{equation}
where $\mathcal{N}(f_{i},f_{j})=\partial f_{i}\partial f_{j}\equiv n^{\mu}\partial_{\mu}f_{i}n^{\nu}\partial_{\nu}f_{j}$. 

\section{Coherent states and NEC}\label{section 4}
Let us consider a generic coherent state given by
\begin{equation}\label{coherent1}
\ket{\mathcal{O}}=\hat{\mathcal{D}}(\{\mathcal{O}\})\ket{0}=\prod_{i}\hat{\mathcal{D}}(\mathcal{O}_{i})\ket{0},
\end{equation}
where $\hat{\mathcal{D}}(\mathcal{O}_{i})=e^{\mathcal{O}_{i}\hat{a}_{i}^{\dagger}-\mathcal{O}_{i}^{*}\hat{a}_{i}}$. Hence, the expectation value of the operator in (\ref{operator}) \textit{w.r.t} the state (\ref{coherent1}) is given by
\begin{equation}
\begin{split}
\bra{\mathcal{O}}:\hat{T}_{\mu\nu}:n^{\mu}n^{\nu}\ket{\mathcal{O}} & =\sum_{i,j}\Big[\mathcal{N}(f_{i},f_{j})\mathcal{O}_{i}\mathcal{O}_{j}+\mathcal{N}(f_{i}^{*},f_{j})\mathcal{O}_{i}^{*}\mathcal{O}_{j}+\mathcal{N}(f_{i},f_{j}^{*})\mathcal{O}_{i}\mathcal{O}_{j}^{*}\\
 & +\mathcal{N}(f_{i}^{*},f_{j}^{*})\mathcal{O}_{i}^{*}\mathcal{O}_{j}^{*}\Big].
\end{split}
\end{equation}
The above expression can further be simplified in the following way
\begin{equation}\label{result1}
\begin{split}
\bra{\mathcal{O}}:\hat{T}_{\mu\nu}:n^{\mu}n^{\nu}\ket{\mathcal{O}} & =\sum_{i,j}\Big[\mathcal{O}_{i}(\mathcal{O}_{j}\mathcal{N}(f_{i},f_{j})+\mathcal{N}(f_{i},f_{j}^{*})\mathcal{O}_{j}^{*})+\mathcal{O}_{i}^{*}(\mathcal{O}_{j}\mathcal{N}(f_{i}^{*},f_{j})\\
+\mathcal{N}(f_{i}^{*},f_{j}^{*})\mathcal{O}_{j}^{*})\Big] & =\sum_{i,j}\mathcal{N}(\mathcal{O}_{i}f_{i}+\mathcal{O}_{i}^{*}f_{i}^{*},\mathcal{O}_{j}f_{j}+\mathcal{O}_{j}^{*}f_{j}^{*})=4\mathcal{N}(\bar{\mathcal{O}},\bar{\mathcal{O}})\geq0, \ \bar{\mathcal{O}}=\sum_{i}\text{Re}(\mathcal{O}_{i}f_{i}).
\end{split}
\end{equation}
The above expression leads to the following theorem.  
\begin{theorem}\label{Theorem1}
The set of all the coherent states $\{\ket{\mathcal{O}}\}$ of the field operator $\hat{\phi}$ holds NEC.
\end{theorem}
Let us now consider two such coherent states given by $\ket{\mathcal{O}^{(1)}}$ and $\ket{\mathcal{O}^{(2)}}$, then
\begin{equation}\label{interference}
\begin{split}
\bra{\mathcal{O}^{(1)}}:\hat{T}_{\mu\nu}:n^{\mu}n^{\nu}\ket{\mathcal{O}^{(2)}} & =\sum_{i,j}\Big[\mathcal{N}(f_{i},f_{j})\mathcal{O}_{i}^{(2)}\mathcal{O}_{j}^{(2)}+\mathcal{N}(f_{i},f_{j}^{*})\mathcal{O}_{i}^{(2)}\mathcal{O}_{j}^{(1)*}\\
 & +\mathcal{N}(f_{i}^{*},f_{j})\mathcal{O}_{i}^{(1)*}\mathcal{O}_{j}^{(2)}+\mathcal{N}(f_{i}^{*},f_{j}^{*})\mathcal{O}_{i}^{(1)*}\mathcal{O}_{j}^{(1)*}\Big]\\
 & =\sum_{i,j}\mathcal{N}(f_{i}\mathcal{O}_{i}^{(2)}+f_{i}^{*}\mathcal{O}_{i}^{(1)*},f_{j}\mathcal{O}_{j}^{(2)}+f_{j}^{*}\mathcal{O}_{j}^{(1)*})\braket{\mathcal{O}^{(1)}|\mathcal{O}^{(2)}}.
\end{split}
\end{equation}
Using the result in (\ref{inner-product}), we can express
\begin{equation}
\braket{\mathcal{O}^{(1)}|\mathcal{O}^{(2)}}=e^{-\frac{1}{2}||\mathcal{O}^{(1)}-\mathcal{O}^{(2)}||^{2}+i\Delta(\mathcal{O}^{(1)},\mathcal{O}^{(2)})}\equiv\prod_{i}\Big[e^{-\frac{1}{2}|\mathcal{O}_{i}^{(1)}-\mathcal{O}_{i}^{(2)}|^{2}+i\text{Im}(\mathcal{O}_{i}^{(1)*}\mathcal{O}_{i}^{(2)})}\Big].
\end{equation}
Hence, the expression (\ref{interference}) can be expressed as
\begin{equation}\label{result2}
\bra{\mathcal{O}^{(1)}}:\hat{T}_{\mu\nu}:n^{\mu}n^{\nu}\ket{\mathcal{O}^{(2)}}=\mathcal{N}(F,F)e^{-\frac{1}{2}||\mathcal{O}^{(1)}-\mathcal{O}^{(2)}||^{2}+i\Delta(\mathcal{O}^{(1)},\mathcal{O}^{(2)})},
\end{equation}
where $F=\sum_{i}(f_{i}\mathcal{O}_{i}^{(2)}+f_{i}^{*}\mathcal{O}_{i}^{(1)*})$ is a complex function. The expression (\ref{result2}) has a huge implication which we discuss now.

In equation (\ref{0.1}), it is shown that the quantum states which do not violate NEC do not necessarily form a vector space. Although the theorem \ref{Theorem1} shows that the set of coherent states satisfy NEC, however, (\ref{result2}) shows that all the combinations of coherent states do not satisfy NEC, as expected. In the next few theorems, we discuss a few possible combinations of coherent states with some constraints that satisfy NEC. This helps in constructing a subset of Hilbert space containing quantum states that satisfy NEC, therefore, the following theorems play an important role in classifying quantum states satisfying NEC.

Given two coherent states $\ket{\mathcal{O}^{(1)}}$ and $\ket{\mathcal{O}^{(2)}}$  holding NEC condition, we obtain a two-dimensional vector space which is nothing but the span of $\ket{\mathcal{O}^{(1)}}$ and $\ket{\mathcal{O}^{(2)}}$. Let us consider a generic state from this two-dimensional vector space, denoted by $\ket{\psi_{+}}=\mathcal{A}(\ket{\mathcal{O}^{(1)}}+e^{\delta_{1}+i\delta_{2}}\ket{\mathcal{O}^{(2)}})$, then
\begin{equation}\label{plus-state}
\begin{split}
\bra{\psi_{+}} & :\hat{T}_{\mu\nu}:n^{\mu}n^{\nu}\ket{\psi_{+}}=|\mathcal{A}|^{2}\Big[4\mathcal{N}(\bar{\mathcal{O}}_{1},\bar{\mathcal{O}}_{1})+4e^{2\delta_{1}}\mathcal{N}(\bar{\mathcal{O}}_{2},\bar{\mathcal{O}}_{2})\\
+\mathcal{N}(F,F) & e^{\delta_{1}-\frac{1}{2}||\mathcal{O}^{(1)}-\mathcal{O}^{(2)}||^{2}+i(\Delta(\mathcal{O}^{(1)},\mathcal{O}^{(2)})+\delta_{2})}+\mathcal{N}(F^{*},F^{*})e^{\delta_{1}-\frac{1}{2}||\mathcal{O}^{(1)}-\mathcal{O}^{(2)}||^{2}-i(\Delta(\mathcal{O}^{(1)},\mathcal{O}^{(2)})+\delta_{2})}\Big]\\
=4|\mathcal{A}|^{2} & \Big[\mathcal{N}(\bar{\mathcal{O}}_{1},\bar{\mathcal{O}}_{1})+e^{2\delta_{1}}\mathcal{N}(\bar{\mathcal{O}}_{2},\bar{\mathcal{O}}_{2})+\frac{e^{\delta_{1}}}{2}\text{Re}(\mathcal{N}(F,F)e^{i(\Delta(\mathcal{O}^{(1)},\mathcal{O}^{(2)})+\delta_{2})})e^{-\frac{1}{2}||\mathcal{O}^{(1)}-\mathcal{O}^{(2)}||^{2}}\Big].
\end{split}
\end{equation}
Similarly, for the state $\ket{\psi_{-}}=\mathcal{A}(\ket{\mathcal{O}^{(1)}}-e^{\delta_{1}+i\delta_{2}}\ket{\mathcal{O}^{(2)}})$,
\begin{equation}\label{minus-state}
\begin{split}
\bra{\psi_{-}} & :\hat{T}_{\mu\nu}:n^{\mu}n^{\nu}\ket{\psi_{-}}=4|\mathcal{A}|^{2}\Big[\mathcal{N}(\bar{\mathcal{O}}_{1},\bar{\mathcal{O}}_{1})+e^{2\delta_{1}}\mathcal{N}(\bar{\mathcal{O}}_{2},\bar{\mathcal{O}}_{2})\\
 & -\frac{e^{\delta_{1}}}{2}\text{Re}(\mathcal{N}(F,F)e^{i(\Delta(\mathcal{O}^{(1)},\mathcal{O}^{(2)})+\delta_{2})})e^{-\frac{1}{2}||\mathcal{O}^{(1)}-\mathcal{O}^{(2)}||^{2}}\Big].
\end{split} 
\end{equation}
The expressions of the expectation values of $:\hat{T}_{\mu\nu}:n^{\mu}n^{\nu}$ \textit{w.r.t} the states $\ket{\psi_{\pm}}$ in (\ref{plus-state}) and (\ref{minus-state}) lead to the following theorem. 
\begin{theorem}\label{Theorem2}
Given two coherent states $\ket{\mathcal{O}^{(1)}}$ and $\ket{\mathcal{O}^{(2)}}$, the states of the form $\ket{\psi_{\pm}}=\mathcal{A}(\ket{\mathcal{O}^{(1)}}\pm e^{\delta_{1}+i\delta_{2}}\ket{\mathcal{O}^{(2)}})$ hold NEC, provided the following condition is satisfied
\begin{equation}\label{inequality}
\begin{split}
\mathcal{N}(\bar{\mathcal{O}}_{1},\bar{\mathcal{O}}_{1}) & +e^{2\delta_{1}}\mathcal{N}(\bar{\mathcal{O}}_{2},\bar{\mathcal{O}}_{2})\pm\frac{e^{\delta_{1}}}{2}\text{Re}(\mathcal{N}(F,F)e^{i(\Delta(\mathcal{O}^{(1)},\mathcal{O}^{(2)})+\delta_{2})})e^{-\frac{1}{2}||\mathcal{O}^{(1)}-\mathcal{O}^{(2)}||^{2}}\geq0\\
\implies e^{\delta_{1}}\Big|\text{Re}[\mathcal{N}(F,F) & e^{i(\Delta(\mathcal{O}^{(1)},\mathcal{O}^{(2)})+\delta_{2})}]\Big|e^{-\frac{1}{2}||\mathcal{O}^{(1)}-\mathcal{O}^{(2)}||^{2}}\leq2[\mathcal{N}(\bar{\mathcal{O}}_{1},\bar{\mathcal{O}}_{1})+e^{2\delta_{1}}\mathcal{N}(\bar{\mathcal{O}}_{2},\bar{\mathcal{O}}_{2})].
\end{split}
\end{equation}
\end{theorem}
Since the function $\mathcal{N}(F,F)$ is a complex function, we can express it as $\mathcal{N}(F(x),F(x))=\mathcal{F}(x)e^{i\tilde{\Delta}_{12}(x)}$ where $\mathcal{F}(x)$ is a positive real-valued function. Hence, the inequality in (\ref{inequality}) can be expressed as follows
\begin{equation}\label{inequality2}
\begin{split}
\mathcal{F}(x)|\cos(\Delta(\mathcal{O}^{(1)},\mathcal{O}^{(2)})+\tilde{\Delta}_{12}(x)+\delta_{2})|e^{-\frac{1}{2}||\mathcal{O}^{(1)}-\mathcal{O}^{(2)}||^{2}} &\leq2[e^{-\delta_{1}}\mathcal{N}(\bar{\mathcal{O}}_{1},\bar{\mathcal{O}}_{1})+e^{\delta_{1}}\mathcal{N}(\bar{\mathcal{O}}_{2},\bar{\mathcal{O}}_{2})]\\
 & =2(e^{-\delta_{1}}\mathcal{F}_{1}(x)+e^{\delta_{1}}\mathcal{F}_{2}(x)).
\end{split}
\end{equation}
where $\mathcal{N}(\bar{\mathcal{O}}_{i},\bar{\mathcal{O}}_{i})=\mathcal{F}_{i}(x)$s are positive real-valued functions. An example of the inequality (\ref{inequality}) is provided in the Appendix. The inequality in (\ref{inequality2}) leads to the following theorem.
\begin{theorem}\label{Theorem3}
If the following condition
\begin{equation}\label{condition}
\frac{e^{-\delta_{1}}\mathcal{F}_{1}(x)+e^{\delta_{1}}\mathcal{F}_{2}(x)}{\mathcal{F}(x)}e^{\frac{1}{2}||\mathcal{O}^{(1)}-\mathcal{O}^{(2)}||^{2}}\geq\frac{1}{2},
\end{equation} 
holds for all the points in the spacetime, then the states of the form $\ket{\psi_{\pm}}$ definitely hold NEC.
\end{theorem}
The above condition strongly depends on the states $\ket{\mathcal{O}_{1}},\ket{\mathcal{O}_{2}}$ and the mode solutions of the Klein-Gordon equation. It is quite clear that the condition (\ref{condition}) is more likely to be satisfied if $||\mathcal{O}^{(1)}-\mathcal{O}^{(2)}||\gg1$ or $|\delta_{1}|\gg1$.  It is important to note that the above theorems are still valid even if we consider the mass term in the action (\ref{action}).

Recall the following definitions
\begin{equation}
\begin{split}
\bar{\mathcal{O}}_{1}=\frac{1}{2}\sum_{i}(\mathcal{O}_{i}^{(1)}f_{i}+\mathcal{O}_{i}^{(1)*}f_{i}^{*}), & \ \bar{\mathcal{O}}_{2}=\frac{1}{2}\sum_{i}(\mathcal{O}_{i}^{(2)}f_{i}+\mathcal{O}_{i}^{(2)*}f_{i}^{*}), \ F=\sum_{i}(f_{i}\mathcal{O}_{i}^{(2)}+f_{i}^{*}\mathcal{O}_{i}^{(1)*}),
\end{split}
\end{equation}
and therefore, we can write the following
\begin{equation}
\begin{split}
\mathcal{N}(\bar{\mathcal{O}}_{1},\bar{\mathcal{O}}_{2}) & =\frac{1}{4}\sum_{i,j}\Big[\mathcal{O}_{i}^{(1)}\mathcal{O}_{j}^{(2)}\partial f_{i}\partial f_{j}+\mathcal{O}_{i}^{(1)}\mathcal{O}_{j}^{(2)*}\partial f_{i}\partial f_{j}^{*}+\mathcal{O}_{i}^{(1)*}\mathcal{O}_{j}^{(2)}\partial f_{i}^{*}\partial f_{j}+\mathcal{O}_{i}^{(1)*}\mathcal{O}_{j}^{(2)*}\partial f_{i}^{*}\partial f_{j}^{*}\Big]\\
|\mathcal{N}(F,F)| & =\partial F\partial F^{*}=\sum_{i,j}\Big[\mathcal{O}_{i}^{(1)}\mathcal{O}_{j}^{(2)}\partial f_{i}\partial f_{j}+\mathcal{O}_{i}^{(1)}\mathcal{O}_{j}^{(1)*}\partial f_{i}\partial f_{j}^{*}+\mathcal{O}_{i}^{(2)*}\mathcal{O}_{j}^{(2)}\partial f_{i}^{*}\partial f_{j}+\mathcal{O}_{i}^{(2)*}\mathcal{O}_{j}^{(1)*}\partial f_{i}^{*}\partial f_{j}^{*}\Big],
\end{split}
\end{equation}
where $\partial\equiv n^{\mu}\partial_{\mu}$. Defining $\mathcal{O}_{f}^{(1,2)}\equiv\sum_{i}f_{i}\mathcal{O}_{i}^{(1,2)}$ and using the above expressions, we obtain the following inequality
\begin{equation}
\begin{split}
|\mathcal{N}(F,F)|-4\mathcal{N}(\bar{\mathcal{O}}_{1},\bar{\mathcal{O}}_{2}) & =\Big[\partial\mathcal{O}_{f}^{(1)}\partial\mathcal{O}_{f}^{(1)*}+\partial\mathcal{O}_{f}^{(2)}\partial\mathcal{O}_{f}^{(2)*}-\partial\mathcal{O}_{f}^{(1)}\partial\mathcal{O}_{f}^{(2)*}-\partial\mathcal{O}_{f}^{(1)*}\partial\mathcal{O}_{f}^{(2)}\Big]\\
 & =\partial(\mathcal{O}_{f}^{(1)}-\mathcal{O}_{f}^{(2)})\partial(\mathcal{O}_{f}^{(1)*}-\mathcal{O}_{f}^{(2)*})=|\partial(\mathcal{O}_{f}^{(1)}-\mathcal{O}_{f}^{(2)})|^{2}\geq0\\
\implies\frac{\mathcal{N}(\bar{\mathcal{O}}_{1},\bar{\mathcal{O}}_{2})}{|\mathcal{N}(F,F)|} & \leq\frac{1}{4}.
\end{split}
\end{equation}
From the theorem (\ref{Theorem2}) and theorem (\ref{Theorem3}), it follows that if $\frac{e^{-\delta_{1}}\mathcal{F}_{1}(x)+e^{\delta_{1}}\mathcal{F}_{2}(x)}{\mathcal{F}(x)}e^{\frac{1}{2}||\mathcal{O}^{(1)}-\mathcal{O}^{(2)}||^{2}}<\frac{1}{2}$, atleast for a spacetime point $x$ and $\delta_{1}$, and 
\begin{equation}\label{NEC-violation}
\mathcal{F}(x)|\cos(\Delta(\mathcal{O}^{(1)},\mathcal{O}^{(2)})+\tilde{\Delta}_{12}(x)+\delta_{2})|e^{-\frac{1}{2}||\mathcal{O}^{(1)}-\mathcal{O}^{(2)}||^{2}}>2[e^{-\delta_{1}}\mathcal{N}(\bar{\mathcal{O}}_{1},\bar{\mathcal{O}}_{1})+e^{\delta_{1}}\mathcal{N}(\bar{\mathcal{O}}_{2},\bar{\mathcal{O}}_{2})],
\end{equation}
then the states $\ket{\psi_{\pm}}$ violates NEC. Since $\frac{e^{-\delta_{1}}\mathcal{F}_{1}(x)+e^{\delta_{1}}\mathcal{F}_{2}(x)}{\mathcal{F}(x)}e^{\frac{1}{2}||\mathcal{O}^{(1)}-\mathcal{O}^{(2)}||^{2}}<\frac{1}{2}$, there always exists a relative phase $\bar{\delta}_{2}$ for which (\ref{NEC-violation}) is satisfied. This shows that all the states belong to the linear span of any two coherent states do not hold NEC. Further, the minimum value of the function $\frac{e^{-\delta_{1}}\mathcal{F}_{1}(x)+e^{\delta_{1}}\mathcal{F}_{2}(x)}{\mathcal{F}(x)}$ is $2\frac{\sqrt{\mathcal{F}_{1}(x)\mathcal{F}_{2}(x)}}{\mathcal{F}(x)}$. Therefore, if there exists a $\bar{\delta}_{2}$ for which the following condition
\begin{equation}
|\cos(\Delta(\mathcal{O}^{(1)},\mathcal{O}^{(2)})+\tilde{\Delta}_{12}(x)+\delta_{2})|e^{-\frac{1}{2}||\mathcal{O}^{(1)}-\mathcal{O}^{(2)}||^{2}}>4\frac{\sqrt{\mathcal{F}_{1}(x)\mathcal{F}_{2}(x)}}{\mathcal{F}(x)},
\end{equation}
is satisfied atleast for a spacetime point $x$ then, NEC is violated. It is important to remember that $\delta_{2}\in[0,\pi]$ for the states $\ket{\psi_{\pm}}$.

The above-mentioned approach can be generalized to a $m$-dimensional subspace of the Hilbert space of matter quantum states. In order to see this, let us consider a collection of $m$ linear independent coherent states $\{\ket{\mathcal{O}^{(i)}}\}_{i=1}^{m}$. A generic state belongs to the linear span of the coherent states $\{\ket{\mathcal{O}^{(i)}}\}_{i=1}^{m}$ can be expressed as follows
\begin{equation}\label{4.17}
\ket{\psi}=\mathcal{A}[\ket{\mathcal{O}^{(1)}}+e^{\delta_{2}+i\phi_{2}}\ket{\mathcal{O}^{(2)}}+e^{\delta_{3}+i\phi_{3}}\ket{\mathcal{O}^{(3)}}+\ldots+e^{\delta_{m}+i\phi_{m}}\ket{\mathcal{O}^{(m)}}],
\end{equation}
where $\{\delta_{i}\}_{i=2}^{m}\in\mathbb{R}$ and $\{\phi_{i}\}_{i=2}^{m}\in[0,2\pi)$. The expectation value of $:\hat{T}_{\mu\nu}:n^{\mu}n^{\nu}$ \textit{w.r.t} the state $\ket{\psi}$ can be expressed as follows
\begin{equation}\label{4.18}
\begin{split}
\bra{\psi}:\hat{T}_{\mu\nu}: & n^{\mu}n^{\nu}\ket{\psi}=4|\mathcal{A}|^{2}\Big[\sum_{i=1}^{m}e^{2\delta_{i}}\mathcal{N}(\bar{\mathcal{O}}_{i},\bar{\mathcal{O}}_{i})+\sum_{i<j}\frac{e^{\delta_{i}+\delta_{j}}}{2}\text{Re}\left(\mathcal{N}(F_{ij},F_{ij})e^{i(\Delta(\mathcal{O}^{(i)},\mathcal{O}^{(j)})+\phi_{i}+\phi_{j})}\right)e^{-\frac{1}{2}||\mathcal{O}^{(i)}-\mathcal{O}^{(j)}||^{2}}\Big]\\
F_{ij}(x) & =\sum_{k}(f_{k}(x)\mathcal{O}_{k}^{(j)}+f_{k}^{*}(x)\mathcal{O}_{k}^{(i)*}), \ \delta_{1}=\phi_{1}=0, \ \Delta(\mathcal{O}^{(i)},\mathcal{O}^{(j)})=\sum_{k}\text{Im}(\mathcal{O}_{k}^{(i)*}\mathcal{O}_{k}^{(j)}). 
\end{split}
\end{equation}
Expressing $\mathcal{N}(F_{ij}(x),F_{ij}(x))=\mathcal{F}_{ij}(x)e^{i\Delta_{ij}(x)}$, the above expression can be re-expressed as follows
\begin{equation}
\begin{split}
\bra{\psi}:\hat{T}_{\mu\nu}(x):n^{\mu}n^{\nu}\ket{\psi} & =4|\mathcal{A}|^{2}\Bigg[\sum_{i=1}^{m}e^{2\delta_{i}}\mathcal{N}(\bar{\mathcal{O}}_{i}(x),\bar{\mathcal{O}}_{i}(x))+\sum_{i<j}\Big[\frac{e^{\delta_{i}+\delta_{j}}\mathcal{F}_{ij}(x)}{2}e^{-\frac{1}{2}||\mathcal{O}^{(i)}-\mathcal{O}^{(j)}||^{2}}\\
 & \times\cos(\Delta(\mathcal{O}^{(i)},\mathcal{O}^{(j)})+\Delta_{ij}(x)+\phi_{i}+\phi_{j})\Big]\Bigg].
\end{split}
\end{equation}
The above expression leads to the following theorem, a generalized version of the theorem (\ref{Theorem2}).
\begin{theorem}\label{Theorem4}
Given a collection of linear independent coherent states $\{\ket{\mathcal{O}^{(i)}}\}_{i=1}^{m}$, the states of the form $\ket{\psi}$ (shown earlier) violate NEC if for any spacetime point $x$ and the sets of parameters $\{\delta_{i}\}_{i=2}^{m}, \{\phi_{i}\}_{i=2}^{m}$ 
\begin{equation}
\sum_{i<j}\frac{e^{\delta_{i}+\delta_{j}}\mathcal{F}_{ij}(x)}{2}\cos(\Delta(\mathcal{O}^{(i)},\mathcal{O}^{(j)})+\Delta_{ij}(x)+\phi_{i}+\phi_{j})e^{-\frac{1}{2}||\mathcal{O}^{(i)}-\mathcal{O}^{(j)}||^{2}}<-\sum_{i=1}^{m}e^{2\delta_{i}}\mathcal{N}(\bar{\mathcal{O}}_{i}(x),\bar{\mathcal{O}}_{i}(x)),
\end{equation}
is satisfied provided
\begin{equation}
\sum_{i=1}^{m}e^{2\delta_{i}}\mathcal{N}(\bar{\mathcal{O}}_{i}(x),\bar{\mathcal{O}}_{i}(x))<\sum_{i<j}\frac{e^{\delta_{i}+\delta_{j}}\mathcal{F}_{ij}(x)}{2}e^{-\frac{1}{2}||\mathcal{O}^{(i)}-\mathcal{O}^{(j)}||^{2}}.
\end{equation}
\end{theorem}
If for any value of $m$, the above condition holds then, we conclude that NEC is violated by the matter. Theorem \ref{Theorem2} and \ref{Theorem4} give the condition in terms of a mathematical inequality under which a linear combination of coherent states of the form given in (\ref{4.17}) violates NEC. On the other hand, the theorem \ref{Theorem3} directly follows from the theorem \ref{Theorem2}.

Suppose a matter in a generic curved spacetime is given by a density matrix \cite{fano1957description, sabbaghzadeh2007role} $\hat{\rho}$ which could be either a pure or a mixed state. Then we obtain the following relation
\begin{equation}
\begin{split}
\langle:\hat{T}_{\mu\nu}:n^{\mu}n^{\nu}\rangle & =\text{Tr}[\hat{\rho}:\hat{T}_{\mu\nu}(x):n^{\mu}n^{\nu}]=\sum_{j}\bra{\mathcal{O}^{(j)}}\hat{\rho}:\hat{T}_{\mu\nu}(x):n^{\mu}n^{\nu}\ket{\mathcal{O}^{(j)}}\\
 & =\sum_{i,j}\bra{\mathcal{O}^{(j)}}\hat{\rho}\ket{\mathcal{O}^{(i)}}\bra{\mathcal{O}^{(i)}}:\hat{T}_{\mu\nu}(x):n^{\mu}n^{\nu}\ket{\mathcal{O}^{(j)}}\\
 & =\sum_{i,j}\Big[\bra{\mathcal{O}^{(j)}}\hat{\rho}\ket{\mathcal{O}^{(i)}}\mathcal{N}(F_{ij},F_{ij})e^{-\frac{1}{2}||\mathcal{O}^{(i)}-\mathcal{O}^{(j)}||^{2}}e^{i\Delta(\mathcal{O}^{(i)},\mathcal{O}^{(j)})}\Big],
\end{split}
\end{equation}
where $\{\ket{\mathcal{O}^{(i)}}\}$ is the set of complete coherent state basis. Defining $\rho_{ji}\equiv\frac{\bra{\mathcal{O}^{(j)}}\hat{\rho}\ket{\mathcal{O}^{(i)}}}{\braket{\mathcal{O}^{(j)}|\mathcal{O}^{(i)}}}$, the above expression can further be simplified as follows
\begin{equation}
\begin{split}
\langle:\hat{T}_{\mu\nu}:n^{\mu}n^{\nu}\rangle & =\sum_{i,j}\Big[\rho_{ji} \ \mathcal{N}(F_{ij},F_{ij})e^{-||\mathcal{O}^{(i)}-\mathcal{O}^{(j)}||^{2}}\Big]\\
 & =\sum_{i,j}\Big[\text{Re}\left(\rho_{ji} \ \mathcal{N}(F_{ij},F_{ij})\right)e^{-||\mathcal{O}^{(i)}-\mathcal{O}^{(j)}||^{2}}\Big],
\end{split}
\end{equation}
where we have used the following relation
\begin{equation}
\Delta(\mathcal{O}^{(i)},\mathcal{O}^{(j)})=-\Delta(\mathcal{O}^{(j)},\mathcal{O}^{(i)}).
\end{equation}
\begin{theorem}
A density matrix $\hat{\rho}$, describing the state of matter in a generic curved spacetime would violate NEC if and only if there exists a spacetime point $x$ such that the following condition
\begin{equation}
\sum_{i,j}\Big[\text{Re}[\rho_{ji} \ \mathcal{N}(F_{ij}(x),F_{ij}(x))]e^{-||\mathcal{O}^{(i)}-\mathcal{O}^{(j)}||^{2}}\Big]<0,
\end{equation}
is satisfied.
\end{theorem}   
   
\section{Dynamical violation of NEC}
Earlier, we discussed that there exist certain combinations of coherent states satisfying NEC. However, it is not quite clear under what circumstances the time-evolution of a NEC satisfying quantum state also satisfies NEC. The result of this section is important in order to find out the existence of spacetime singularity from the Raychaudhuri equation which becomes an evolution equation when the affine parameter is chosen to be the coordinate time. Moreover, in quantum theory, the nature of a solution of the Raychaudhuri equation depends on the many-body matter quantum state, and in particular, it depends on the null energy condition of that state, expressed by the quantity $\bra{\psi(t)}:\hat{T}_{\mu\nu}(t,\vec{x}):n^{\mu}(t,\vec{x})n^{\nu}(t,\vec{x})\ket{\psi(t)}$ in the interaction picture. In this section, it is shown that the time-evolution can indeed map a NEC satisfying quantum state outside the set of NEC satisfying quantum states. The possibility of this kind of violation of NEC depends on the Hamiltonian of the system, the underlying curved spacetime, and the quantum state at the initial time $\ket{\psi(t_{0})}$. 
 
\subsection{Mathematical formulation}
In this section, we address whether a quantum state of matter can violate NEC dynamically or not i.e. can the time evolution of a quantum state of matter satisfying NEC, generate a state that violates NEC. In order to compute the time evolution of quantum states and observables in an interacting QFT, the interaction picture is often used. In order to define the interaction picture, the Hamiltonian is divided into two parts $\hat{H}=\hat{H}_{0}+\hat{H}_{\text{int}}$. In this picture, the operators carry the time dependence through the free Hamiltonian $\hat{H}_{0}$ whereas the states carry the time dependence through the interaction Hamiltonian $\hat{H}_{\text{int}}$ in the following way
\begin{equation}\label{6.1}
i\frac{d}{dt}\ket{\psi(t)}_{I}=\hat{H}_{\text{int},I}\ket{\psi(t)}_{I}, \ \hat{\mathcal{O}}_{I}(t)=e^{i\hat{H}_{0}t}\hat{\mathcal{O}}_{S}e^{-i\hat{H}_{0}t}
\end{equation}
where $\hat{\mathcal{O}}_{S}$ is the operator in the Schr$\ddot{o}$dinger picture and $\hat{\mathcal{O}}_{I}(t)$ is the operator in the interaction picture. Now onwards, we drop the labels $I$ and $S$.

Let us consider a state $\ket{\psi(t_{1})}$ satisfying $\bra{\psi(t_{1})}:\hat{T}_{\mu\nu}(t_{1},\vec{x}):n^{\mu}(t_{1},\vec{x})n^{\nu}(t_{1},\vec{x})\ket{\psi(t_{1})}\geq0$. Then, at time $t_{2}>t_{1}$, the quantum state becomes 
\begin{equation}
\ket{\psi(t_{2})}=\mathcal{T}\left(e^{-i\int_{t_{1}}^{t_{2}}\hat{H}_{\text{int}}(t)dt}\right)\ket{\psi(t_{1})},
\end{equation}
where $\mathcal{T}$ is the time-ordering operator. Using the above time-evolved state, we obtain the following relation 
\begin{equation}\label{6.3}
\begin{split}
\bra{\psi(t_{2})} & :\hat{T}_{\mu\nu}(t_{2},\vec{x}):n^{\mu}(t_{2},\vec{x})n^{\nu}(t_{2},\vec{x})\ket{\psi(t_{2})}\\
 & =\bra{\psi(t_{1})}\bar{\mathcal{T}}\left(e^{i\int_{t_{1}}^{t_{2}}\hat{H}_{\text{int}}(t)dt}\right):\hat{T}_{\mu\nu}(t_{2},\vec{x}):\mathcal{T}\left(e^{-i\int_{t_{1}}^{t_{2}}\hat{H}_{\text{int}}(t)dt}\right)\ket{\psi(t_{1})}n^{\mu}(t_{2},\vec{x})n^{\nu}(t_{2},\vec{x})\\
 & \equiv\bra{\psi(t_{1})}\tilde{\mathcal{T}}\left(e^{i\int_{t_{1}}^{t_{2}}\hat{H}_{\text{int}}(t)dt}:\hat{T}_{\mu\nu}(t_{2},\vec{x}):e^{-i\int_{t_{1}}^{t_{2}}\hat{H}_{\text{int}}(t)dt}\right)\ket{\psi(t_{1})}n^{\mu}(t_{2},\vec{x})n^{\nu}(t_{2},\vec{x}),
\end{split}
\end{equation}  
where $\bar{\mathcal{T}}$ is the anti time-ordering operator. If $\hat{H}_{0}(t)$ is time-independent, then $\mathcal{T}\left(e^{-i\int_{t_{1}}^{t_{2}}\hat{H}_{0}(t)dt}\right)=e^{-i\hat{H}_{0}(t_{2}-t_{1})}$. For the sake of simplicity, we defined the operation $\tilde{\mathcal{T}}$ in the last line of the equation (\ref{6.3}).   

Further, the exponent in time-evolution operator can be expressed in terms of the Hamiltonian density $\int_{t_{1}}^{t_{2}}\hat{H}_{\text{int}}(t)dt=\int d^{4}y \ \hat{\mathcal{H}}_{\text{int}}(y)$ where $\hat{\mathcal{H}}(y)$ is the Hamiltonian density operator. Hence, we obtain the following relation 
\begin{equation}\label{commutation}
\begin{split}
\tilde{\mathcal{T}} & \left(e^{i\int_{t_{1}}^{t_{2}}\hat{H}_{\text{int}}(t)dt}:\hat{T}_{\mu\nu}(t_{2},\vec{x}):e^{-i\int_{t_{1}}^{t_{2}}\hat{H}_{\text{int}}(t)dt}\right)=\bar{\mathcal{T}}\left(e^{i\int_{t_{1}}^{t_{2}}\hat{H}_{\text{int}}(t)dt}\right):\hat{T}_{\mu\nu}(t_{2},\vec{x}):\mathcal{T}\left(e^{-i\int_{t_{1}}^{t_{2}}\hat{H}_{\text{int}}(t)dt}\right)\\
 & =:\hat{T}_{\mu\nu} (t_{2},\vec{x}):+i\int d^{4}y[\hat{\mathcal{H}}_{\text{int}}(y),:\hat{T}_{\mu\nu}(t_{2},\vec{x}):]+\ldots.
\end{split}
\end{equation}
The above relation shows that the violation of NEC by a quantum state under time evolution depends on the commutator of the second term of the above equation in the leading order. The above commutator takes into account the effect of curved spacetime, as the Hamiltonian density and the energy-momentum tensor both depend on the metric explicitly for a minimally coupled field theory. From the micro-causality condition, it follows that $[\hat{\mathcal{O}}_{1}(x),\hat{\mathcal{O}}_{2}(y)]=0$ if and only if the spacetime points $x$ and $y$ are spacelike separated. As a consequence of this relation, only the operators located within the past light-cone of the spacetime point $(t_{2},\vec{x})$ give rise to non-zero commutators in (\ref{commutation}). 

Analogously, in the case of a system described by a mixed state with the density matrix $\hat{\rho}$, we obtain the following relation
\begin{equation}\label{6.7}
\begin{split}
\langle:\hat{T}_{\mu\nu} & (t_{2},\vec{x}):\rangle n^{\mu}(t_{2},\vec{x})n^{\nu}(t_{2},\vec{x})=\text{Tr}[\hat{\rho}(t_{2}):\hat{T}_{\mu\nu}(t_{2},\vec{x}):]n^{\mu}(t_{2},\vec{x})n^{\nu}(t_{2},\vec{x})\\
 & =\text{Tr}\Bigg[\mathcal{T}\left(e^{-i\int_{t_{1}}^{t_{2}}\hat{H}_{\text{int}}(t)dt}\right)\hat{\rho}(t_{1})\bar{\mathcal{T}}\left(e^{i\int_{t_{1}}^{t_{2}}\hat{H}_{\text{int}}(t)dt}\right):\hat{T}_{\mu\nu}(t_{2},\vec{x}):\Bigg]n^{\mu}(t_{2},\vec{x})n^{\nu}(t_{2},\vec{x})\\
 & =\text{Tr}\Bigg[\hat{\rho}(t_{1})\tilde{\mathcal{T}}\left(e^{i\int_{t_{1}}^{t_{2}}\hat{H}_{\text{int}}(t)dt}:\hat{T}_{\mu\nu}(t_{2},\vec{x}): \ e^{-i\int_{t_{1}}^{t_{2}}\hat{H}_{\text{int}}(t)dt}\right)\Bigg]n^{\mu}(t_{2},\vec{x})n^{\nu}(t_{2},\vec{x}),
\end{split}
\end{equation} 
where the cyclicity property of trace operation is used. Hence, the results (\ref{6.3}) and (\ref{6.7}) are almost the same. If we consider the density matrix to be a pure state ($\hat{\rho}(t_{1})=\ket{\psi(t_{1})}\bra{\psi(t_{1})}$), then the expression in (\ref{6.7}) reduces to the expression in (\ref{6.3}), however, the expression in (\ref{6.7}) is also valid for a mixed state. Therefore, the expression in (\ref{6.7}) is much more general than the expression in (\ref{6.3}) in describing the null energy condition of a time-evolved quantum state.

\subsection{Massless $\phi^{3}$ scalar field theory in a static spacetime}
Let us consider a massless scalar field theory in a static curved spacetime with the $\phi^{3}$-interaction. The corresponding action is given by
\begin{equation}\label{interacting action}
\mathcal{S}=-\int\sqrt{-g} \ d^{4}x\Big[\frac{1}{2}g^{\mu\nu}\partial_{\mu}\phi\partial_{\nu}\phi+V(\phi)\Big], \ V(\phi)=\frac{\lambda}{3!}\phi^{3}.
\end{equation}
The conjugate momentum variable is given by
\begin{equation}
\Pi(x)=-\sqrt{-g(x)}g^{0\nu}\partial_{\nu}\phi(x)=-\sqrt{-g(x)}\partial^{0}\phi(x),
\end{equation}
and the only non-zero equal-time commutation bracket is
\begin{equation}
[\hat{\phi}(x),\hat{\Pi}(y)]=i\delta^{(3)}(\vec{x}-\vec{y})\implies[\hat{\phi}(x),\partial^{0}\hat{\phi}(y)]=-i\frac{\delta^{(3)}(\vec{x}-\vec{y})}{\sqrt{-g(x)}}, \ x^{0}=y^{0}.
\end{equation}  
Hence, the Hamiltonian density is given by
\begin{equation}
\begin{split}
\mathcal{H}(x) & =-\sqrt{-g(x)}\partial^{0}\phi(x)\partial_{0}\phi(x)+\sqrt{-g(x)}\Big[\frac{1}{2}g^{\mu\nu}(x)\partial_{\mu}\phi(x)\partial_{\nu}\phi(x)+V(\phi(x))\Big]\\
 & =\sqrt{-g(x)}\Big[-\frac{1}{2}\partial_{0}\phi(x)\partial^{0}\phi(x)+\frac{1}{2}\partial_{i}\phi(x)\partial^{i}\phi(x)+V(\phi(x))\Big].
\end{split}
\end{equation}
In a static spacetime, the above expression reduces to the following
\begin{equation}
\mathcal{H}(x)=\sqrt{-g(x)}\Big[\frac{1}{2g(x)g^{00}(x)}\Pi^{2}(x)+\frac{1}{2}\partial_{i}\phi(x)\partial^{i}\phi(x)+V(\phi(x))\Big].
\end{equation}
In this section, we discuss the effect of $V(\phi)=\frac{\lambda}{3!}\phi^{3}$ interaction through the perturbative technique on the violation of NEC by quantum states under time evolution. In order to apply the perturbative technique, we follow the interaction picture of quantum field theory \cite{Schwartz:2013pla}. Hence, the field operator can be expanded on a basis consists of mode solutions of free d'Alembert's equation in the following manner
\begin{equation}
\hat{\phi}(x)=\sum_{i}[\hat{a}_{i}g_{i}(x)+\hat{a}_{i}^{\dagger}g_{i}^{*}(x)],
\end{equation} 
where $\Box g_{i}(x)=0$, and the creation and annihilation operators satisfy the algebra $[\hat{a}_{i},\hat{a}_{j}^{\dagger}]=\delta_{ij}$. As a result of this expansion, the general commutation bracket is given by
\begin{equation}
[\hat{\phi}(x),\hat{\phi}(y)]=\Delta_{\phi}^{(0)}(x,y), \ [\hat{\Pi}(x),\hat{\Pi}(y)]=\Delta_{\Pi}^{(0)}(x,y), \ [\hat{\phi}(x),\hat{\Pi}(y)]=\Delta^{(0)}(x,y),
\end{equation}
where
\begin{equation}
\begin{split}
\Delta_{\phi}^{(0)}(x,y) & =\sum_{i}[g_{i}(x)g_{i}^{*}(y)-g_{i}^{*}(x)g_{i}(y)]=2i\sum_{i}\text{Im}(g_{i}(x)g_{i}^{*}(y))\\
\Delta_{\Pi}^{(0)}(x,y) & =\sqrt{g(x)g(y)}\partial_{x}^{0}\partial_{y}^{0}\Delta_{\phi}^{(0)}(x,y), \ \Delta^{(0)}(x,y)=-\sqrt{-g(y)}\partial_{y}^{0}\Delta_{\phi}^{(0)}(x,y).
\end{split}
\end{equation}
On the other hand, we also obtain the following relation
\begin{equation}\label{6.16}
\partial_{\mu}(n^{\mu}(x)n^{\nu}(x)\partial_{\nu}\hat{\phi}(x))=\sum_{i}[\hat{a}_{i}\bar{\Delta}_{g_{i}}(x)+\hat{a}_{i}^{\dagger}\bar{\Delta}_{g_{i}^{*}}(x)]=\sum_{i}[\hat{a}_{i}\bar{\Delta}_{g_{i}}(x)+\hat{a}_{i}^{\dagger}\bar{\Delta}_{g_{i}}^{*}(x)],
\end{equation}
where
\begin{equation}
\bar{\Delta}_{g_{i}}(x)=\partial_{\mu}(n^{\mu}(x)n^{\nu}(x)\partial_{\nu}g_{i}(x)), \ \bar{\Delta}_{g_{i}^{*}}(x)=\partial_{\mu}(n^{\mu}(x)n^{\nu}(x)\partial_{\nu}g_{i}^{*}(x))=\bar{\Delta}_{g_{i}}^{*}(x).
\end{equation}
Since $:\hat{T}_{\mu\nu}(x):n^{\mu}(x)n^{\nu}(x)=:(\partial\hat{\phi}(x))^{2}:$, we obtain the following commutation relation
\begin{equation}\label{6.18}
[\hat{\mathcal{H}}_{\text{int}}(y),:\hat{T}_{\mu\nu}(x):n^{\mu}(x)n^{\nu}(x)]=2\sqrt{-g(y)}\Delta_{\phi}(x,y)\frac{\delta V(\phi)}{\delta\phi(y)}\partial_{\mu}(n^{\mu}(x)n^{\nu}(x)\partial_{\nu}\hat{\phi}(x)).
\end{equation}
Considering $V(\phi)=\frac{\lambda}{3!}\phi^{3}$, we obtain the following relation
\begin{equation}
\frac{\delta V(\phi)}{\delta\phi}=\frac{\lambda}{2}\phi^{2}(y)=\sum_{i,j}\Big[\mathcal{V}_{ij}(y)\hat{a}_{i}\hat{a}_{j}+\mathcal{V}_{\bar{i}j}(y)\hat{a}_{i}^{\dagger}\hat{a}_{j}+\mathcal{V}_{i\bar{j}}(y)\hat{a}_{i}\hat{a}_{j}^{\dagger}+\mathcal{V}_{\bar{i}\bar{j}}(y)\hat{a}_{i}^{\dagger}\hat{a}_{j}^{\dagger}\Big],
\end{equation}
where
\begin{equation}
\begin{split}
\mathcal{V}_{ij}(y) & =\frac{\lambda}{2}g_{i}(y)g_{j}(y), \ \mathcal{V}_{\bar{i}j}(y)=\frac{\lambda}{2}g_{i}^{*}(y)g_{j}(y)\\ 
\mathcal{V}_{i\bar{j}}(y) & =\frac{\lambda}{2}g_{i}(y)g_{j}^{*}(y), \ \mathcal{V}_{\bar{i}\bar{j}}(y)=\frac{\lambda}{2}g_{i}^{*}(y)g_{j}^{*}(y).
\end{split}
\end{equation}
As a result of the above relations, we can write
\begin{equation}\label{6.21}
\begin{split}
n^{\mu}(x)n^{\nu}(x)\tilde{\mathcal{T}} & \left(e^{i\int_{t_{1}}^{t_{2}}\hat{H}_{\text{int}}(t)dt}:\hat{T}_{\mu\nu}(t_{2},\vec{x}):e^{-i\int_{t_{1}}^{t_{2}}\hat{H}_{\text{int}}(t)dt}\right)=:\hat{T}_{\mu\nu}(t_{2},\vec{x}):n^{\mu}(x)n^{\nu}(x)\\
+2i\partial_{\mu} & (n^{\mu}(x)n^{\nu}(x)\partial_{\nu}\hat{\phi}(x))\int d^{4}y\sqrt{-g(y)}\Big[\Delta_{\phi}^{(0)}(x,y)\frac{\delta V(\phi)}{\delta\phi(y)}\Big]-\ldots,
\end{split}
\end{equation}
where $\hat{H}_{\text{int}}(t)$ denotes the interaction Hamiltonian in the interaction picture which in this case is given by $\int\sqrt{-g(x)}d^{3}x\frac{\lambda}{3!}\phi^{3}(x)$. The integral in the above expression can further be simplified as follows
\begin{equation}
2i\int d^{4}y\sqrt{-g(y)}\Delta_{\phi}^{(0)}(x,y)\frac{\delta V(\phi)}{\delta\phi(y)}=i\sum_{i,j}\Big[\bar{\mathcal{V}}_{ij}(x)\hat{a}_{i}\hat{a}_{j}+\bar{\mathcal{V}}_{\bar{i}j}(x)\hat{a}_{i}^{\dagger}\hat{a}_{j}+\bar{\mathcal{V}}_{i\bar{j}}(x)\hat{a}_{i}\hat{a}_{j}^{\dagger}+\bar{\mathcal{V}}_{\bar{i}\bar{j}}(x)\hat{a}_{i}^{\dagger}\hat{a}_{j}^{\dagger}\Big],
\end{equation} 
where  
\begin{equation}
\begin{split}
\bar{\mathcal{V}}_{ij}(x) & =\lambda\int d^{4}y\sqrt{-g(y)}\Delta_{\phi}^{(0)}(x,y)g_{i}(y)g_{j}(y), \ \bar{\mathcal{V}}_{\bar{i}j}(x)=\lambda\int d^{4}y\sqrt{-g(y)}\Delta_{\phi}^{(0)}(x,y)g_{i}^{*}(y)g_{j}(y)\\
\bar{\mathcal{V}}_{i\bar{j}}(x) & =\lambda\int d^{4}y\sqrt{-g(y)}\Delta_{\phi}^{(0)}(x,y)g_{i}(y)g_{j}^{*}(y), \ \bar{\mathcal{V}}_{\bar{i}\bar{j}}(x)=\lambda\int d^{4}y\sqrt{-g(y)}\Delta_{\phi}^{(0)}(x,y)g_{i}^{*}(y)g_{j}^{*}(y). 
\end{split}
\end{equation}
Hence, the second term in the equation (\ref{6.21}) can be expressed as
\begin{equation}\label{6.24}
\begin{split}
i & \sum_{k}[\hat{a}_{k}\bar{\Delta}_{g_{k}}(x)+\hat{a}_{k}^{\dagger}\bar{\Delta}_{g_{k}}^{*}(x)]\sum_{i,j}\Big[\bar{\mathcal{V}}_{ij}(x)\hat{a}_{i}\hat{a}_{j}+\bar{\mathcal{V}}_{\bar{i}j}(x)\hat{a}_{i}^{\dagger}\hat{a}_{j}+\bar{\mathcal{V}}_{i\bar{j}}(x)\hat{a}_{i}\hat{a}_{j}^{\dagger}+\bar{\mathcal{V}}_{\bar{i}\bar{j}}(x)\hat{a}_{i}^{\dagger}\hat{a}_{j}^{\dagger}\Big]\\
=i & \sum_{i,j,k}\Bigg[\bar{\Delta}_{g_{k}}(x)\Big[\bar{\mathcal{V}}_{ij}(x)\hat{a}_{i}\hat{a}_{j}\hat{a}_{k}+\bar{\mathcal{V}}_{\bar{i}j}(x)\hat{a}_{i}^{\dagger}\hat{a}_{j}\hat{a}_{k}+\bar{\mathcal{V}}_{i\bar{j}}(x)\hat{a}_{j}^{\dagger}\hat{a}_{i}\hat{a}_{k}+\bar{\mathcal{V}}_{\bar{i}\bar{j}}(x)\hat{a}_{i}^{\dagger}\hat{a}_{j}^{\dagger}\hat{a}_{k}\Big]\\
 & +\bar{\Delta}_{g_{k}}^{*}(x)\Big[\bar{\mathcal{V}}_{ij}(x)\hat{a}_{k}^{\dagger}\hat{a}_{i}\hat{a}_{j}+\bar{\mathcal{V}}_{\bar{i}j}(x)\hat{a}_{i}^{\dagger}\hat{a}_{k}^{\dagger}\hat{a}_{j}+\bar{\mathcal{V}}_{i\bar{j}}(x)\hat{a}_{j}^{\dagger}\hat{a}_{k}^{\dagger}\hat{a}_{i}+\bar{\mathcal{V}}_{\bar{i}\bar{j}}(x)\hat{a}_{i}^{\dagger}\hat{a}_{j}^{\dagger}\hat{a}_{k}^{\dagger}\Big]\Bigg]\\
+i & \Big[2\sum_{i,j}\bar{\Delta}_{g_{i}}(x)(\bar{\mathcal{V}}_{\bar{i}j}(x)\hat{a}_{j}+\bar{\mathcal{V}}_{\bar{i}\bar{j}}(x)\hat{a}_{j}^{\dagger})+\sum_{i,k}\bar{\mathcal{V}}_{i\bar{i}}(x)(\hat{a}_{k}\bar{\Delta}_{g_{k}}(x)+\bar{\Delta}_{g_{k}}^{*}(x)\hat{a}_{k}^{\dagger})\Big]. 
\end{split}
\end{equation}
The above equation clearly shows that the coherent states or eigenstates of the annihilation operators will not remain the eigenstates of the annihilation operators under time evolution. Further, the interaction term can in principle lead to the
violation of NEC by the quantum states under time evolution, which is argued below. This also shows that the violation of NEC depends on the functions $\{\bar{\mathcal{V}}_{ij}(x),\ldots,\bar{\mathcal{V}}_{\bar{i}\bar{j}}(x)\}$ consist of mode solutions, hence, they are directly connected to the geometry. Further, this violation also depends on the properties of the quantum states since it depends on the action of the creation and annihilation operators on the quantum states.

The terms in the equation (\ref{6.24}) affect the NEC condition of the time-evolved states significantly in the leading order. If this expression sandwiched between $\bra{\psi(t_{1})}$ and $\ket{\psi(t_{1})}$ is negative, and its magnitude is more than $\bra{\psi(t_{1})}:\hat{T}_{\mu\nu}(t_{2},\vec{x}):n^{\mu}(t_{2},\vec{x})n^{\nu}(t_{2},\vec{x})\ket{\psi(t_{1})}$, then the $\bra{\psi(t_{2})}:\hat{T}_{\mu\nu}(t_{2},\vec{x}):n^{\mu}(t_{2},\vec{x})n^{\nu}(t_{2},\vec{x})\ket{\psi(t_{2})}<0$ in the leading order quantum correction. The expressions in (\ref{4.18}) and (\ref{6.24}) are made out of the solutions of the Klein-Gordon equations. Since the term in the first parenthesis of (\ref{6.24}) sandwiched between $\bra{\psi(t_{1})}$ and $\ket{\psi(t_{1})}$ and the expression (\ref{4.18}) are cubic and quadratic in $\{\mathcal{O}_{i}^{(k)}\}_{k=1}^{m}$ respectively, hence, considering a state such that $\{|\mathcal{O}_{i}^{(k)}|\rightarrow0\}_{k=1}^{m}$, we can effectively neglect the effect of these terms compared to the terms in the second parenthesis of (\ref{6.24}) sandwiched between $\bra{\psi(t_{1})}$ and $\ket{\psi(t_{1})}$. On the other hand, there exist spacetimes \cite{Ford:1997hb, jacobson2005introduction, parker2009quantum, alsing2001phase, al2018dirac, lehn2018klein} in which the mode solutions of the Klein-Gordon equation are oscillating in nature. As a result, we can always find a spacetime point in this class of spacetimes at which $\bra{\psi(t_{2})}:\hat{T}_{\mu\nu}(t_{2},\vec{x}):n^{\mu}(t_{2},\vec{x})n^{\nu}(t_{2},\vec{x})\ket{\psi(t_{2})}$ becomes negative in the leading order quantum correction due to the oscillating nature of the terms in the second parenthesis (\ref{6.24}). Moreover, as $\{\mathcal{O}_{i}^{(k)}\}_{k=1}^{m}$ are the state-dependent free parameters, it is possible to choose such parameters or in other words, construct a quantum state suitably which can violate NEC after time evolution. This clearly shows that the time-evolution can map a NEC satisfying quantum state to a NEC violating quantum state.

\section{Average null energy condition}
In this section, we show the dependence of ANEC on the geometry of spacetime and the quantum states of matter. In order to show that, we rewrite the mathematical inequality of ANEC in a different manner. As shown earlier, the ANEC \textit{w.r.t} a quantum state $\ket{\psi}$ mathematically demands
\begin{equation}\label{7.1}
\int_{\gamma}\langle T_{\mu\nu}(X(\lambda))\rangle l^{\mu}(\lambda)l^{\nu}(\lambda)d\lambda\geq0,
\end{equation}
where $l^{\mu}(\lambda)=\frac{dX^{\mu}(\lambda)}{d\lambda}$ is the tangent vector of the null geodesic $\gamma$ and $\langle T_{\mu\nu}(X(\lambda))\rangle=\bra{\psi}T_{\mu\nu}(X(\lambda))\ket{\psi}$. The above expression can also be expressed as follows
\begin{equation}\label{ANEC1}
\begin{split}
\int_{\gamma} & \langle T_{\mu\nu}(X(\lambda))\rangle l^{\mu}(\lambda)l^{\nu}(\lambda)d\lambda=\int d^{4}x\langle T_{\mu\nu}(x)\rangle\int\frac{d^{4}k}{(2\pi)^{4}}\int_{\gamma}e^{ik.(x-X(\lambda))}\dot{X}^{\mu}(\lambda)\dot{X}^{\nu}(\lambda)d\lambda\\
 & =\int\frac{d^{4}k}{(2\pi)^{4}}\langle T_{\mu\nu}(-k)\rangle\int_{\gamma} e^{-ik.X(\lambda)}\dot{X}^{\mu}(\lambda)\dot{X}^{\nu}(\lambda)d\lambda,
\end{split}
\end{equation}
where `dot' represents derivative \textit{w.r.t} $\lambda$. We make progress further by using the following result
\begin{equation}\label{ANEC2}
\begin{split}
\int_{\gamma} e^{-ik.X(\lambda)} & \dot{X}^{\mu}(\lambda)\dot{X}^{\nu}(\lambda)d\lambda=\int_{\gamma}\frac{i\dot{X}^{\mu}(\lambda)\dot{X}^{\nu}(\lambda)}{k.\dot{X}(\lambda)}\frac{d}{d\lambda}e^{-ik.X(\lambda)}\\
 & =i\Big[\frac{\dot{X}_{2}^{\mu}\dot{X}_{2}^{\nu}}{k.\dot{X}_{2}}e^{-ik.X_{2}}-\frac{\dot{X}_{1}^{\mu}\dot{X}_{1}^{\nu}}{k.\dot{X}_{1}}e^{-ik.X_{1}}\Big]-i\int_{\gamma}\frac{d}{d\lambda}\left(\frac{\dot{X}^{\mu}(\lambda)\dot{X}^{\nu}(\lambda)}{k.\dot{X}(\lambda)}\right)e^{-ik.X(\lambda)}d\lambda,
\end{split}
\end{equation}
where $\dot{X}(\lambda_{1,2})=\dot{X}_{1,2}$ and $X(\lambda_{1,2})=X_{1,2}$. In the above expression, we obtain a boundary and a bulk term. Plugging (\ref{ANEC2}) in (\ref{ANEC1}), we obtain the following expression
\begin{equation}
\begin{split}
\int_{\gamma} & \langle T_{\mu\nu}(X(\lambda))\rangle l^{\mu}(\lambda)l^{\nu}(\lambda)d\lambda=i\int\frac{d^{4}k}{(2\pi)^{4}}\langle T_{\mu\nu}(-k)\rangle\Big[\frac{\dot{X}_{2}^{\mu}\dot{X}_{2}^{\nu}}{k.\dot{X}_{2}}e^{-ik.X_{2}}-\frac{\dot{X}_{1}^{\mu}\dot{X}_{1}^{\nu}}{k.\dot{X}_{1}}e^{-ik.X_{1}}\Big]\\
 & -i\int\frac{d^{4}k}{(2\pi)^{4}}\langle T_{\mu\nu}(-k)\rangle\int_{\gamma}\frac{e^{-ik.X(\lambda)}}{(k.\dot{X}(\lambda))^{2}}k_{\rho}[2\ddot{X}^{(\mu}(\lambda)\dot{X}^{\nu)}(\lambda)\dot{X}^{\rho}-\ddot{X}^{\rho}\dot{X}^{\mu}\dot{X}^{\nu}]d\lambda.
\end{split}
\end{equation}
Using the geodesic equations, the above expression reduces to
\begin{equation}\label{7.5}
\begin{split}
\int_{\gamma} & \langle T_{\mu\nu}(X(\lambda))\rangle l^{\mu}(\lambda)l^{\nu}(\lambda)d\lambda=i\int\frac{d^{4}k}{(2\pi)^{4}}\langle T_{\mu\nu}(-k)\rangle\Big[\frac{\dot{X}_{2}^{\mu}\dot{X}_{2}^{\nu}}{k.\dot{X}_{2}}e^{-ik.X_{2}}-\frac{\dot{X}_{1}^{\mu}\dot{X}_{1}^{\nu}}{k.\dot{X}_{1}}e^{-ik.X_{1}}\Big]\\
 & +i\int\frac{d^{4}k}{(2\pi)^{4}}\langle T_{\mu\nu}(-k)\rangle\int_{\gamma}\Big[\frac{e^{-ik.X(\lambda)}}{(k.\dot{X}(\lambda))^{2}}k_{\rho}[2\Gamma_{ \ \sigma_{1}\sigma_{2}}^{(\mu}(X(\lambda))\dot{X}^{\nu)}(\lambda)\dot{X}^{\rho}(\lambda)\\
 & -\Gamma_{ \ \sigma_{1}\sigma_{2}}^{\rho}(X(\lambda))\dot{X}^{\mu}(\lambda)\dot{X}^{\nu}(\lambda)]\dot{X}^{\sigma_{1}}(\lambda)\dot{X}^{\sigma_{2}}(\lambda)\Big]d\lambda.
\end{split}
\end{equation}
The first term in the above expression is boundary term as it depends on the end-points of the null geodesic $\gamma$ and the second term is the bulk terms as it depends on the nature of the null geodesic $\gamma$. However, both the terms depend on the quantum state $\ket{\psi}$ through $\langle T_{\mu\nu}(-k)\rangle$. Hence, for an arbitrary stress-energy tensor, ANEC demands
\begin{equation}\label{7.6}
\begin{split}
\text{Im}\Bigg[\int\frac{d^{4}k}{(2\pi)^{4}} & \langle T_{\mu\nu}(-k)\rangle\Big[\frac{\dot{X}_{2}^{\mu}\dot{X}_{2}^{\nu}}{k.\dot{X}_{2}}e^{-ik.X_{2}}-\frac{\dot{X}_{1}^{\mu}\dot{X}_{1}^{\nu}}{k.\dot{X}_{1}}e^{-ik.X_{1}}\Big]\Bigg]\leq-\text{Im}\Bigg[\int\frac{d^{4}k}{(2\pi)^{4}}\langle T_{\mu\nu}(-k)\rangle\\
\times\int_{\gamma}\frac{e^{-ik.X(\lambda)}}{(k.\dot{X}(\lambda))^{2}}k_{\rho} & \Big[[2\Gamma_{ \ \sigma_{1}\sigma_{2}}^{(\mu}(X(\lambda))\dot{X}^{\nu)}(\lambda)\dot{X}^{\rho}(\lambda)-\Gamma_{ \ \sigma_{1}\sigma_{2}}^{\rho}(X(\lambda))\dot{X}^{\mu}(\lambda)\dot{X}^{\nu}(\lambda)]\dot{X}^{\sigma_{1}}(\lambda)\dot{X}^{\sigma_{2}}(\lambda)\Big]d\lambda\Bigg].
\end{split}
\end{equation}
On the other hand, since $\int_{\gamma}\langle T_{\mu\nu}(X(\lambda))\rangle l^{\mu}(\lambda)l^{\nu}(\lambda)d\lambda$ is a real quantity, we expect the following equality
\begin{equation}\label{7.7}
\begin{split}
\text{Re}\Bigg[\int\frac{d^{4}k}{(2\pi)^{4}} & \langle T_{\mu\nu}(-k)\rangle\Big[\frac{\dot{X}_{2}^{\mu}\dot{X}_{2}^{\nu}}{k.\dot{X}_{2}}e^{-ik.X_{2}}-\frac{\dot{X}_{1}^{\mu}\dot{X}_{1}^{\nu}}{k.\dot{X}_{1}}e^{-ik.X_{1}}\Big]\Bigg]=-\text{Re}\Bigg[\int\frac{d^{4}k}{(2\pi)^{4}}\langle T_{\mu\nu}(-k)\rangle\\
\times\int_{\gamma}\frac{e^{-ik.X(\lambda)}}{(k.\dot{X}(\lambda))^{2}}k_{\rho} & \Big[[2\Gamma_{ \ \sigma_{1}\sigma_{2}}^{(\mu}(X(\lambda))\dot{X}^{\nu)}(\lambda)\dot{X}^{\rho}(\lambda)-\Gamma_{ \ \sigma_{1}\sigma_{2}}^{\rho}(X(\lambda))\dot{X}^{\mu}(\lambda)\dot{X}^{\nu}(\lambda)]\dot{X}^{\sigma_{1}}(\lambda)\dot{X}^{\sigma_{2}}(\lambda)\Big]d\lambda\Bigg].
\end{split}
\end{equation}
For an interacting scalar field theory given by (\ref{interacting action}), we only need the following contribution
\begin{equation}
\langle T_{\mu\nu}(-k)\rangle=\int\frac{d^{4}q}{(2\pi)^{4}}\langle\phi(q)\phi(-k-q)\rangle q_{\mu}(k_{\nu}+q_{\nu}).
\end{equation}
Therefore, in this case, we obtain the following relations
\begin{equation}\label{7.9}
\begin{split}
\int\frac{d^{4}k}{(2\pi)^{4}}\langle T_{\mu\nu}(-k)\rangle & \Big[\frac{\dot{X}_{2}^{\mu}\dot{X}_{2}^{\nu}}{k.\dot{X}_{2}}e^{-ik.X_{2}}-\frac{\dot{X}_{1}^{\mu}\dot{X}_{1}^{\nu}}{k.\dot{X}_{1}}e^{-ik.X_{1}}\Big]=\int\frac{d^{4}k}{(2\pi)^{4}}\int\frac{d^{4}q}{(2\pi)^{4}}\langle\phi(q)\phi(-k-q)\rangle\\
\times\Bigg[\Big[\left(\frac{(q.\dot{X}_{2})^{2}}{k.\dot{X}_{2}}\right) & e^{-ik.X_{2}}-\left(\frac{(q.\dot{X}_{1})^{2}}{k.\dot{X}_{1}}\right)e^{-ik.X_{1}}\Big]+[q.\dot{X}_{2}e^{-ik.X_{2}}-q.\dot{X}_{1}e^{-ik.X_{1}}]\Bigg],
\end{split}
\end{equation}
and
\begin{equation}\label{7.10}
\begin{split}
\int\frac{d^{4}k}{(2\pi)^{4}}\langle T_{\mu\nu}(-k)\rangle & \int_{\gamma}\frac{e^{-ik.X(\lambda)}}{(k.\dot{X}(\lambda))^{2}}k_{\rho}[2\Gamma_{ \ \sigma_{1}\sigma_{2}}^{(\mu}(X(\lambda))\dot{X}^{\nu)}(\lambda)\dot{X}^{\rho}(\lambda)\\
 & -\Gamma_{ \ \sigma_{1}\sigma_{2}}^{\rho}(X(\lambda))\dot{X}^{\mu}(\lambda)\dot{X}^{\nu}(\lambda)]\dot{X}^{\sigma_{1}}(\lambda)\dot{X}^{\sigma_{2}}(\lambda)d\lambda\\
=\int\frac{d^{4}k}{(2\pi)^{4}}\int\frac{d^{4}q}{(2\pi)^{4}} & \langle\phi(q)\phi(-k-q)\rangle\int_{\gamma} e^{-ik.X(\lambda)}\Big[2\frac{q.\ddot{X}(\lambda)q.\dot{X}(\lambda)}{k.\dot{X}(\lambda)}+q.\ddot{X}(\lambda)\\
 & -\frac{k.\ddot{X}(\lambda)(q.\dot{X}(\lambda))^{2}}{[k.\dot{X}(\lambda)]^{2}}\Big]d\lambda.
\end{split}
\end{equation}
The results in (\ref{7.5}), (\ref{7.6}) and (\ref{7.7}) classify ANEC satisfying quantum states for a given null geodesic since the expressions in these equations depend on the chosen null geodesic and the quantum state through the expectation value of the stress-energy tensor. This also shows apparently that if the null geodesic between two points in spacetime is not unique (existence and uniqueness of a solution of the geodesic equation are valid locally), then the quantum states satisfying ANEC for a null geodesic $\gamma_{1}$ may not satisfy ANEC for another null geodesic $\gamma_{2}$ between the same two boundary points in general. Therefore, a quantum state does not satisfy (\ref{7.1}) in general for all the null curves. As a result, one can classify the quantum states satisfying ANEC for a given null geodesic in that case. Moreover, the equations (\ref{7.9}) and (\ref{7.10}) suggest that for an interacting scalar field theory with polynomial interaction, the state dependence of ANEC comes only from the two-point function of the scalar field \textit{w.r.t} a quantum state and this two-point function also depends on the nature of the interaction. We also want to highlight that for non-unique null geodesics, the dependence of null geodesics in (\ref{7.5}) comes from the projection of momentum modes along the tangent vector of null geodesics and its derivative \textit{w.r.t} the affine parameter. 

\section{Discussion}   
Violation of NEC by quantum states of matter in curved spacetimes put a restriction on the possible quantum states of matter which lead to the stable configuration (see \cite{dubovsky2006null, buniy2006null, buniy2006instabilities}). It is also shown in \cite{dubovsky2006null} that even if a violation of NEC occurs in stable quantum matter, it leads to modes with superluminal propagation. This violates causality. Further, dynamical violation of NEC is another important aspect in cosmological models, discussed in \cite{Vikman:2007sj}.   

Here, we discuss the violation of NEC by the quantum states of matter in a generic curved spacetime using the coherent states of field operators. Further, certain criteria are also discussed by providing some important theorems under which a quantum matter state can hold NEC. These lead to the classification of NEC violating quantum states of matter. This classification helps in avoiding NEC violating quantum states which lead to violation of causality and instabilities. Furthermore, we also show the classification of quantum states based on ANEC. In order to do that, we rewrite ANEC inequality differently in which geometrical and state dependences are shown explicitly.    

\section{Acknowledgement}
SM wants to thank IISER Kolkata for supporting this work through a doctoral fellowship.

\bibliographystyle{unsrt}
\bibliography{draft}

\section{Appendix}
We consider the FRW spacetime in the conformal time coordinate $\eta$. The metric in this coordinate is given by $g_{\mu\nu}=a^{2}(\eta)\eta_{\mu\nu}$ where $\eta_{\mu\nu}=\text{diag}(-1,1,1,1)$. The massive Klein-Gordon equation corresponding to a free scalar field theory, minimally coupled to this spacetime is given by
\begin{equation}
-\frac{1}{\sqrt{-g}}\partial_{\mu}(\sqrt{-g}g^{\mu\nu}\partial_{\nu}\phi)+m^{2}\phi=0.
\end{equation}
The mode solutions of the above equation are given by
\begin{equation}
f_{\vec{k}}(\eta,\vec{x})=\frac{e^{i\vec{k}.\vec{x}-i\int^{\eta}W_{\vec{k}}(\eta')d\eta'}}{a(\eta)\sqrt{2(2\pi)^{3}W_{\vec{k}}(\eta)}},
\end{equation}
where $W_{\vec{k}}(\eta)$ is the solution of the following differential equation
\begin{equation}
W_{\vec{k}}^{2}=\omega_{\vec{k}}^{2}-\frac{\ddot{a}}{a}+\frac{3}{4}\frac{\dot{W}_{\vec{k}}^{2}}{W_{\vec{k}}^{2}}-\frac{1}{2}\frac{\ddot{W}_{\vec{k}}}{W_{\vec{k}}},
\end{equation}
$\omega_{\vec{k}}^{2}(\eta)=\vec{k}^{2}+m^{2}a^{2}(\eta)$, and `dot' represents derivative \textit{w.r.t} the conformal time $\eta$. In the leading order adiabatic approximation \cite{Birrell:1982ix, winitzki2005cosmological, kaya2011stress}, we can write $W_{\vec{k}}(\eta)=\omega_{\vec{k}}(\eta)$.

Now we consider two states $\ket{\psi}_{\pm}=\mathcal{A}[\ket{\mathcal{O}^{(1)}}\pm e^{\delta_{1}+i\delta_{2}}\ket{\mathcal{O}^{(2)}}]$ where $\mathcal{O}_{\vec{k}}^{(1)}$ is non-zero only for $\vec{k}=\vec{k}_{1}$ and $\mathcal{O}_{\vec{k}}^{(2)}$ is non-zero only for $\vec{k}=\vec{k}_{2}\neq\vec{k}_{1}$. Then following the definitions and notations in section \ref{section 4}, we obtain the following expressions
\vfill
\begin{equation}
\begin{split}
\bar{\mathcal{O}}^{(i)}(x) & =\frac{|\mathcal{O}_{\vec{k}_{i}}^{(i)}|}{a(\eta)\sqrt{2(2\pi)^{3}\omega_{\vec{k}_{i}}^{(2)}(\eta)}}\cos\left(\vec{k}_{i}.\vec{x}-\int^{\eta}\omega_{\vec{k}_{i}}(\eta')d\eta'+\Delta_{i}\right),
\end{split}
\end{equation}
considering $\mathcal{O}_{\vec{k}_{1}}^{(1)}=|\mathcal{O}_{\vec{k}_{1}}^{(1)}|e^{i\Delta_{1}}$ and $\mathcal{O}_{\vec{k}_{2}}^{(2)}=|\mathcal{O}_{\vec{k}_{2}}^{(2)}|e^{i\Delta_{2}}$. We consider a null vector $n^{\mu}=\frac{1}{a(\eta)}(1,\vec{\tilde{n}})$ such that $\vec{\tilde{n}}^{2}=1$. Since the time derivatives are small especially for the higher frequency modes which we can choose, we obtain the following expression
\begin{equation}
\mathcal{N}(\bar{\mathcal{O}}^{(i)},\bar{\mathcal{O}}^{(i)})=(\omega_{\vec{k}_{i}}(\eta)-\vec{k}_{i}.\vec{\tilde{n}})^{2}\frac{|\mathcal{O}_{\vec{k}_{i}}^{(i)}|^{2}}{2a^{4}(\eta)(2\pi)^{3}\omega_{\vec{k}_{i}}(\eta)}\sin^{2}\left(\vec{k}_{i}.\vec{x}-\int^{\eta}\omega_{\vec{k}_{i}}(\eta')d\eta'+\Delta_{i}\right).
\end{equation}
Defining $\varphi_{i}(x)\equiv\vec{k}_{i}.\vec{x}-\int^{\eta}\omega_{\vec{k}_{i}}(\eta')d\eta'+\Delta_{i}$, $A_{i}(\vec{k}_{i},\eta)\equiv\frac{|\mathcal{O}_{\vec{k}_{i}}^{(i)}|^{2}}{\omega_{\vec{k}_{i}}(\eta)}$, we obtain the following expression
\begin{equation}
\begin{split}
\text{Re} & [\mathcal{N}(F,F)e^{i\delta_{2}}]=-\frac{1}{2a^{4}(\eta)(2\pi)^{3}}\Big[A_{1}(\vec{k},\eta)\cos(2\varphi_{1}(x)-\delta_{2})(\omega_{\vec{k}_{1}}(\eta)-\vec{k}_{1}.\vec{\tilde{n}})^{2}\\
+A_{2} & (\vec{k}_{2},\eta)\cos(2\varphi_{2}(x)+\delta_{2})(\omega_{\vec{k}_{2}}(\eta)-\vec{k}_{2}.\vec{\tilde{n}})^{2}-2\sqrt{A_{1}(\vec{k}_{1},\eta)A_{2}(\vec{k}_{2},\eta)}\cos(\varphi_{2}(x)-\varphi_{1}(x)+\delta_{2})\\
 & \times(\omega_{\vec{k}_{2}}(\eta)-\vec{k}_{2}.\vec{\tilde{n}})(\omega_{\vec{k}_{1}}(\eta)-\vec{k}_{1}.\vec{\tilde{n}})\Big].   
\end{split}
\end{equation}
Therefore, according to the theorem \ref{Theorem2}, the sates $\ket{\psi}_{\pm}$ satisfy NEC provided the following condition is satisfied
\begin{equation}
\begin{split}
\Bigg|\Bigg[ & A_{1}(\vec{k},\eta)\cos(2\varphi_{1}(x)-\delta_{2})(\omega_{\vec{k}_{1}}(\eta)-\vec{k}_{1}.\vec{\tilde{n}})^{2}+A_{2}(\vec{k}_{2},\eta)\cos(2\varphi_{2}(x)+\delta_{2})(\omega_{\vec{k}_{2}}(\eta)-\vec{k}_{2}.\vec{\tilde{n}})^{2}\\
-2 & \sqrt{A_{1}(\vec{k}_{1},\eta)A_{2}(\vec{k}_{2},\eta)}\cos(\varphi_{2}(x)-\varphi_{1}(x)+\delta_{2})(\omega_{\vec{k}_{2}}(\eta)-\vec{k}_{2}.\vec{\tilde{n}})(\omega_{\vec{k}_{1}}(\eta)-\vec{k}_{1}.\vec{\tilde{n}})\Bigg]\Bigg|e^{-\frac{1}{2}[|\mathcal{O}_{\vec{k}_{1}}^{(1)}|^{2}+|\mathcal{O}_{\vec{k}_{2}}^{(2)}|^{2}]}\\
\leq 2\Big[ & e^{\delta_{1}}(\omega_{\vec{k}_{2}}(\eta)-\vec{k}_{2}.\vec{\tilde{n}})^{2}A_{2}(\vec{k}_{2},\eta)\sin^{2}(\varphi_{2}(x))+e^{-\delta_{1}}(\omega_{\vec{k}_{1}}(\eta)-\vec{k}_{1}.\vec{\tilde{n}})^{2}A_{1}(\vec{k}_{1},\eta)\sin^{2}(\varphi_{1}(x))\Big],
\end{split}
\end{equation}
where we used $||\mathcal{O}^{(1)}-\mathcal{O}^{(2)}||^{2}=|\mathcal{O}_{\vec{k}_{1}}^{(1)}|^{2}+|\mathcal{O}_{\vec{k}_{2}}^{(2)}|^{2}$ in this case. For a given spacetime point $x$, we choose $\delta_{2}=\frac{\pi}{2}-\varphi_{2}(x)+\varphi_{1}(x)$. Then the above inequality becomes
\begin{equation}
\begin{split}
\Bigg| & A_{1}(\vec{k},\eta)\sin(\varphi_{1}(x)+\varphi_{2}(x))(\omega_{\vec{k}_{1}}(\eta)-\vec{k}_{1}.\vec{\tilde{n}})^{2}-A_{2}(\vec{k}_{2},\eta)\sin(\varphi_{2}(x)+\varphi_{1}(x))(\omega_{\vec{k}_{2}}(\eta)-\vec{k}_{2}.\vec{\tilde{n}})^{2}\Bigg|\\
\leq 2 & \Big[e^{\delta_{1}}(\omega_{\vec{k}_{2}}(\eta)-\vec{k}_{2}.\vec{\tilde{n}})^{2}A_{2}(\vec{k}_{2},\eta)\sin^{2}(\varphi_{2}(x))+e^{-\delta_{1}}(\omega_{\vec{k}_{1}}(\eta)-\vec{k}_{1}.\vec{\tilde{n}})^{2}A_{1}(\vec{k}_{1},\eta)\sin^{2}(\varphi_{1}(x))\Big]\\
 & \times e^{\frac{1}{2}[|\mathcal{O}_{\vec{k}_{1}}^{(1)}|^{2}+|\mathcal{O}_{\vec{k}_{2}}^{(2)}|^{2}]}.
\end{split}
\end{equation}
In this situation, if we choose $\delta_{1}\gg1$ and $|\mathcal{O}_{\vec{k}_{2}}^{(2)}|^{2}\ll1$, then the above inequality may not satisfy. As a result, states $\ket{\psi}_{\pm}$ in this condition can violate NEC.

\end{document}